\documentclass[oldversion]{aa}%
\usepackage[pdftex]{color,xcolor}
\usepackage[pdftex]{graphicx}
\usepackage{natbib}
\bibpunct{(}{)}{;}{a}{}{,}
\usepackage{hyperref} 
\hypersetup{pdfauthor=Damien S\'egransan}
\hypersetup{backref=true, pagebackref=true, hyperindex=true, breaklinks=true, colorlinks=true,
 urlcolor=blue, linkcolor=black,  citecolor=blue,
 pagecolor=red, bookmarks=true, bookmarksopen=true}
\usepackage{txfonts}
\begin{document}
  \title{The  CORALIE survey for southern extrasolar planets. }
  \subtitle{XVI. Discovery of a planetary system around   HD~147018 and 
             of two long period and massive planets orbiting  HD~171238 and  HD~204313\thanks{Based on observations 
   collected at the ESO La Silla Observatory with the {\footnotesize CORALIE} Echelle spectrograph mounted on the 
   Swiss telescope.}}

   \author{D. S\'egransan\inst{1}
       \and S.~Udry\inst{1}
        \and M.~Mayor\inst{1} 
       \and D.~Naef\inst{1} 
        \and F.~Pepe\inst{1} 
        \and D.~Queloz\inst{1} 
        \and N.C.~Santos\inst{2,1} 
        \and B-O.~Demory\inst{1}   
        \and P.~Figueira\inst{1}  
            \and M.~Gillon\inst{3,1}      
        \and M.~Marmier\inst{1}  
        \and D.~M\'egevand\inst{1}  
         \and D.~Sosnowska\inst{4}  
         \and O.~Tamuz\inst{1} 
          \and A.~H.M.J.~Triaud\inst{1} 
      } 

   \offprints{Damien S\'egransan, \email{Damien.Segransan@unige.ch}}

   \institute{Observatoire astronomique de l'Universit\'e de Gen\`eve, 
                   51 ch. des Maillettes - Sauverny -, CH-1290 Versoix, 
                   Switzerland
          \and
          Centro de Astrof\'{\i}sica da  Universidade do Porto, Rua das Estrelas, 4150-762 Porto, Portugal
  \and
                   Institut d'Astrophysique et de G\'eophysique, Universit\'e de Li\`ege, 17 All\'ee du Six Ao\^ut, 4000 Li\`ege, Belgium
	         \and
         Laboratoire d'astrophysique, Ecole Polytechnique F\'ed\'erale de Lausanne (EPFL),
         Observatoire de Sauverny, CH-1290 Versoix, Switzerland
	     }

   \date{Received / Accepted}
 \abstract
  { We report the detection of a double planetary system around \object{ {\footnotesize HD}\,140718} as well as 
    the discovery of two long period and  massive planets orbiting  \object{ {\footnotesize HD}\,171238}  
    and   \object{ {\footnotesize HD}\,204313}. Those discoveries were made  with the {\footnotesize CORALIE} Echelle spectrograph 
    mounted on the 1.2-m Euler Swiss telescope located at La Silla Observatory, Chile.
    The planetary system orbiting the nearby G9 dwarf \object{ {\footnotesize HD}\,147018} is composed of an eccentric inner planet  (e=0.47) 
    with twice the mass of Jupiter (2.1~M$_{\rm Jup}$) 
    and  with an orbital period of 44.24~days. The outer planet is even more massive  (6.6~M$_{\rm Jup}$) with a slightly eccentric orbit (e=0.13) 
    and a period of 1008~days.
    The  planet orbiting   \object{ {\footnotesize HD}\,171238} has a minimum mass of  2.6~M$_{\rm Jup}$,  
    a period of 1523~days and an eccentricity of 0.40. It orbits a G8 dwarfs at 2.5~AU.  
    The last planet,\object{ {\footnotesize HD}\,204313}~b, is a 4.0~M$_{\rm Jup}$-planet  
    with a period of 5.3~years and has a low eccentricity ($e=0.13$). It orbits a G5 dwarfs at 3.1~AU.  
    The three parent stars are metal rich, which further strengthened the case that massive planets tend to form 
    around metal rich stars. }
   
 \keywords{
 stars: planetary systems --
 stars: binaries: visual -- 
 techniques: radial velocities --
 stars: individual: \object{ {\footnotesize HD}\,147018}--
 stars: individual: \object{ {\footnotesize HD}\,171238}--
 stars: individual: \object{ {\footnotesize HD}\,204313}}

   \maketitle
\section{Introduction}
   The {\footnotesize CORALIE} radial velocity planet-search program 
   has been ongoing for more 
   than 10 years (start date:June 1998) at the 1.2-meter Swiss telescope 
   located at La Silla Observatory, Chile. 
    It is a  volume limited  planet search survey that contains all Hipparcos
main sequence stars from F8 down to K0 within 50~pc and has a color-dependant distance limit for later type 
stars down to M0  \citep{Udry-2000:a}. The fainter targets do not exceed V=10.
Among the  1647 stars surveyed with {\footnotesize CORALIE}, 40 percent of them are measured with a radial velocity accuracy of 6~ms$^{-1}$ or better and 90 percent of the sample is monitored with an accuracy better than 10~ms$^{-1}$. The remaining 10 percent of the sample have measured at a lower accuracy due to the lower signal to noise ratio and/or to the the widening of the cross-correlation function induced by large stellar rotation velocities.
 {\footnotesize CORALIE} went through a major hardware upgrade in June 2007 to increase the overall efficiency whith an net gain in magnitude of 1.5 which allow to survey the fainter part of the sample with an accuracy of 5-6 ms$^{-1}$.

So far, {\footnotesize CORALIE} has allowed the detection (or has contributed to the detection) of 54 extra-solar planet candidates. This substantial contribution together with discoveries from various other programmes have provided a sample - as of today - of close to 350 exoplanets that now permit to point out interesting statistical constraints for the planet formation and evolution scenarios \cite[see e.g. ][and references therein for reviews on different aspects of the orbital-element distributions or primary star properties]{Ida-2004:b,Marcy-2005, Udry-2007:a, Udry-2007:b,Mordasini-2008}.

Concerning the upper part of the planetary mass distribution, it should be noted that a wealth of massive planets have been and is beeing discovered. For instance, 142 of the 350 discovered planets are more massive than 1~M$_{\rm Jup}$ with periods larger than 50 days.
Their eccentricity distribution is similar to the binaries and only 18 of them are in multiple giant planet systems.
Most of their parent stars are metal rich pointing toward the existence of disk enriched in heavy elements.
Interesting enough,  the massive planet around the metal deficient
star \object{ {\footnotesize HD}\,33636} turns out to be a very low mass stars as shown by FGS/HST astrometric observations
presented in \citep{Bean-2007}. Although not statistically significant, this result show how important astrometric observations 
are to clarify the nature of those massive planets.

 In this paper we report the discovery of a planetary system composed of two massive planets around  \object{ {\footnotesize HD}\,147018} and of two additional long period and massive planets  in orbit around  \object{ {\footnotesize HD}\,171238}, and \object{ {\footnotesize HD}\,204313}. Three of those planets come in addition to the 54 known planets with periods longer than 1000~days and masses larger than 1~M$_{\rm Jup}$.
The paper is organized as follows. In the second section, we discuss the host stars properties. The third section describes the instrumental upgrade of  {\footnotesize CORALIE},  the resulting radial-velocity  measurements and the orbital solutions. In section~\ref{sec:discussion}, 
we provide some concluding remarks. 

\section{Stellar characteristics}
\label{sec:characteristics}   

Effective temperatures, gravities and  metallicities  are derived using the spectroscopic analysis of
 \citet{Santos-2000:b} while the $v.\sin{(i)}$ is computed using \citet{Santos-2002}'s calibration of  {\footnotesize CORALIE}'s Cross-Correlation Function (CCF).
We also used the improved  Hipparcos astrometric  parallaxes re-derived by \citet{vanLeeuwen-2007} to determine the 
the V-band magnitude using the apparent visual magnitude from Hipparcos \citep{ESA-1997}.\\
Metallicities, together with the effective temperatures and absolute V-band magnitudes are used to estimate basic stellar parameters (ages, masses, radii and log g) using theoretical isochrones from \cite{Girardi-2000} and a 
Bayesian estimation method described in \cite{daSilva-2006}. The web interface for the Bayesian estimation of stellar parameters, called 
{\footnotesize PARAM 1.0} can be found at  \href{http://stev.oapd.inaf.it/cgi-bin/param}{http://stev.oapd.inaf.it/cgi-bin/param}.
Resulting Stellar parameters are listed in Table~\ref{table:stellarparameters}.

 \subsection {{\footnotesize HD}\,147018 (HIP~80250)}
\object{ {\footnotesize HD}\,147018} is a G9 dwarf with an astrometric parallax of  $\pi=23.28\pm 0.86$~mas
and an apparent V band magnitude of V$=8.30$. Our spectral analysis results in an effective temperature of T$_{\rm eff}=5441\pm55$~K and a stellar metallicity of $[Fe/H] = 0.10 \pm 0.07$.  
Using theoretical isochrones, we finally derived a mass of $M_{\star}=0.927\pm0.031~M_{\odot}$ with an age of  6.36$\pm$4.33~Gyr. 

 \subsection{ {\footnotesize HD}\,171238 (HIP~91085)}

\begin{figure}[th!]
   \includegraphics[angle=0,width=0.45\textwidth,origin=br]{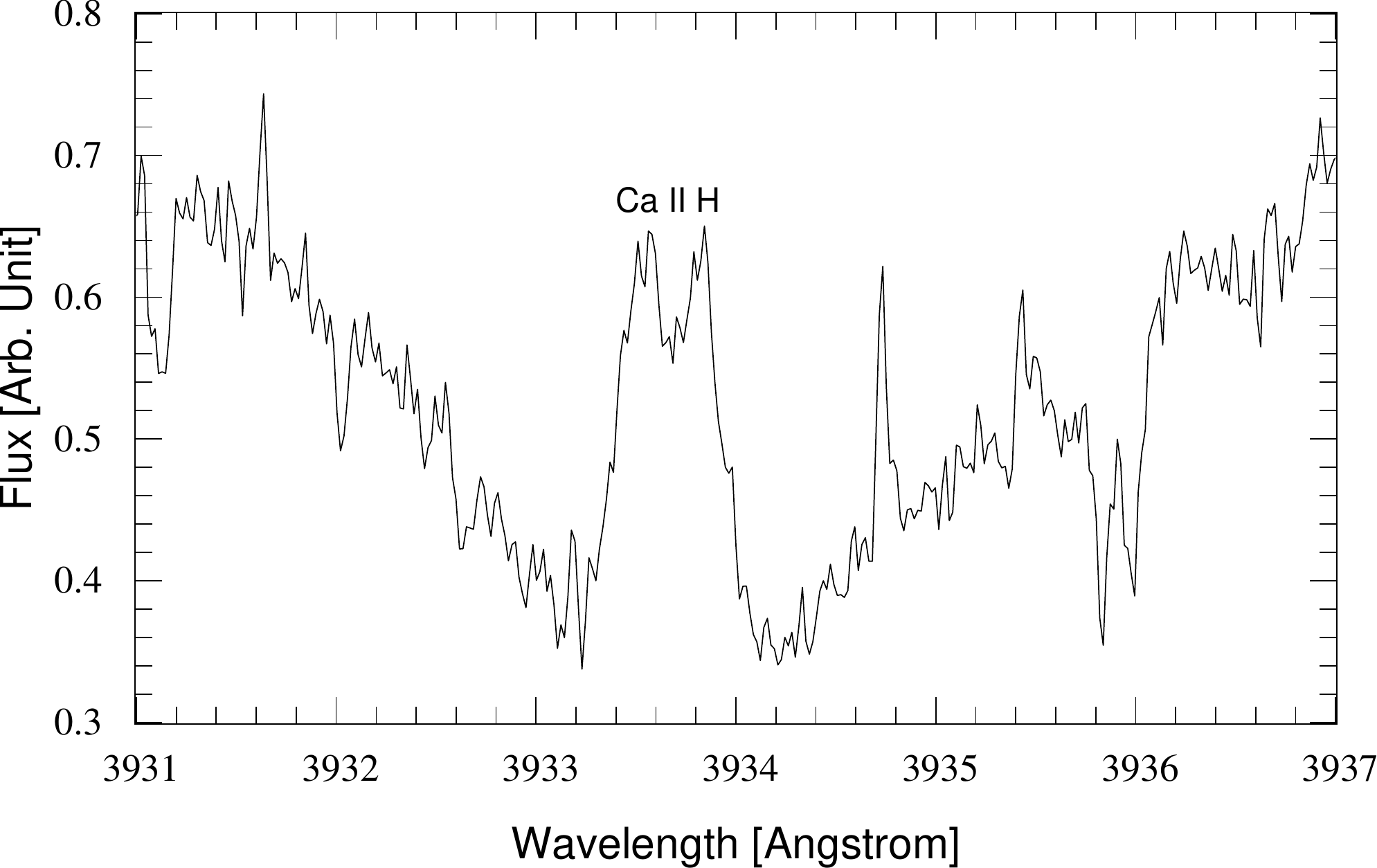} 

\caption[]{
\label{fig:CaIIH-hd171238}
Ca II H emission region for {\footnotesize HD}\,171238. The large re-emission at the bottom of the Ca II H 
absoption line  at $\lambda=3933.66~\AA$ is an indicator of chromospheric activity. This region is contaminated
 by a  few thin thorium emission lines. Thorium pollution is too important to display the Ca II K 
 region (around $\lambda= 3968.47~\AA$).}
\end{figure}

{\footnotesize HD}\,171238 is a G8 dwarf with an astrometric parallax of  $\pi=19.89\pm 1.15$~mas 
and an apparent V band magnitude of V$=8.61$. Our spectral analysis results in an effective temperature of 
T$_{\rm eff}=5467\pm55$~K and a stellar metallicity of $[Fe/H] = 0.17 \pm 0.07$.  
Using theoretical isochrones, we derive a mass of $M_{\star}=0.943\pm0.033~M_{\odot}$ with an age of  4.92$\pm$4.11~Gyr. \\
Eventhough Hipparcos photometry is relatively stable, with a scatter of 0.018~mag in the visible, it should be noted that the star is  listed  in 
the General Catalogue of Variable Stars \citep{Samus-2009} as a BY Draconis-type variable. 
 Such variable stars present photometric variability - induced by spot coverage or by chromospheric activity - up to 0.5 magnitude 
 in the visible on time scales ranging 
 from a fraction of a day to 120 days that are likely to affect the velocities. 
 We were not able to derive a  value of the $\log{\left(R^{'}_{HK}\right)}$ index  since the star is  too faint to conduct a proper spectral
 analysis with { \footnotesize CORALIE}, but we were able to averaged 35 { \footnotesize CORALIE} spectra taken with the Thorium
lamp. As  shown on Fig.\ref{fig:CaIIH-hd171238}, a clear Ca II re-emission is seen at  $\lambda=3933.66~\AA$ revealing the presence
of a significant chromospheric activity possibly induced by stellar spots or plagues.

 \subsection{{\footnotesize HD}\,204313 (HIP~106006)}

{\footnotesize HD}\,204313 is a G5 dwarf with an astrometric parallax of  $\pi = 21.11\pm 0.62$~mas  
and an apparent V band magnitude of V$=7.99$. The spectral analysis results in an effective temperature of 
T$_{\rm eff}=5767\pm17$~K and a stellar metallicity of $[Fe/H] = 0.18 \pm 0.02$.
 Using theoretical isochrones, we  derive a mass of $M_{\star}=1.045\pm0.033~M_{\odot}$ with an age of  3.38$\pm$2.58~Gyr. 

\begin{table}[h!]
\caption{
\label{table:stellarparameters}
Observed and inferred stellar parameters of the 
planets host's stars presented in this paper. (1) : Parameter derived from  \citet{Girardi-2000} models. (2) : Parameter derived using { \footnotesize CORALIE CCF}.}
\begin{center}
\begin{tabular}{llccc}
\hline\hline
Parameters                         &        &\object{ {\footnotesize HD}\,147018}          &\object{ {\footnotesize HD}\,171238}         & \object{ {\footnotesize HD}\,204313}\\
\hline                                                        
Sp. T.                         &            &G9V               &K0V              &  G5V       \\
V                                 &         &8.3               &8.66             &  7.99       \\
$B-V$                        &              &0.763             &0.74             &  0.697                  \\
$\pi$                           & [mas]     &23.28$\pm$0.86    &19.89$\pm$1.15   &  21.11$\pm$0.62\\
 $M_{V}$                   &                &5.13              &5.15             &  4.61 \\
 $T_{eff}$                  & [K]           &5441$\pm$55    &5467$\pm$55      &  5767$\pm$17\\
$\log{g}$                   & [cgs]         &4.38$\pm$0.16                 &4.39$\pm$0.14    &  4.37$\pm$0.05\\
$[Fe/H]$                    & [dex]         &0.10$\pm$0.05              &0.17$\pm$0.07    &  0.18$\pm$0.02              \\
$v \sin{(i)}^{(2)}$                & [kms$^{-1}$]       &1.56              &1.48             &   1.59             \\
$M_{\star}^{(1)}$               &[M$_{\odot}$]      &0.927$\pm$0.031         & 0.943$\pm$0.033            &  1.045$\pm$0.033                  \\
$\log{g}^{(1)}$                   & [cgs]         &4.42$\pm$0.04                &4.43$\pm$0.04    &  4.36$\pm$0.04\\
Age$^{(1)}$&[Gyr]& 6.36$\pm$4.33&4.92$\pm$4.11&3.38$\pm$2.58\\
\hline
\end{tabular}
\end{center}
\end{table}

\section{Radial velocities and orbital solutions \label{sec:measurements}}

\subsection{ {\footnotesize CORALIE} upgrade}
\label{sec:upgrade} 
Triggered by the interest to carry out spectroscopic follow-up on transit candidates
 fainter than the stars surveyed in our main planet search programme (V$<$10), we decided to 
 to improve the overall efficiency of the instrument by upgrading  {\footnotesize CORALIE}  in June 2007.
The fibre link and the cross-disperser optics have been removed and
replaced by a new design. The double scrambler has also been removed and
the grism/prism cross-disperser component
replaced by a series of 4 Schott F2 prisms of 32~$\deg$ angle each. The net outcome of this new design is to maintain the spectral range from 381 to
681 nm but with a large efficiency gain of about
a factor of 6 (8 below 420 nm) and a spectroscopic
resolution of 55 000-60 000 (increased by
10-20\%).  
Those hardware modifications have, however, affected the instrumental zero point
with radial velocity offsets that could reach up to $\approx 20$~ms$^{-1}$, depending on the target spectral type.
For this reason, we decides to refer to the orginal {\footnotesize CORALIE}  as {\footnotesize CORALIE-98} and to the upgraded one as   {\footnotesize CORALIE-07}. The overall instrumental precision was not affected by the upgrate and stays at the  5~ms$^{-1}$ level. 

The direct consequence - on our main planet search survey -  of the upgrade to {\footnotesize CORALIE-07} is to
 to increase the efficiency of the instrument on bright targets (V$<$8.5) and to improve the radial velocity 
 accuracy on the fainter part of our sample (28\% of the sample, V=8.5-10).
Those stars are now monitored with a long term radial velocity accuracy of 5-6~ms$^{-1}$.\\

 \subsection{Two massive planets in orbit around {\footnotesize HD}\,147018}
 \object{ {\footnotesize HD}\,147018}  has been observed with {\footnotesize CORALIE} at La Silla 
Obervatory since May 2003. Six radial-velocity  measurements with a typical signal-to-noise ratio of 25 
(per pixel at 550 nm) were obtained with   {\footnotesize CORALIE-98} leading to a mean measurement uncertainty of 6.1~ms$^{-1}$, including photon noise and calibration errors. An additional 105 radial-velocity  measurements  were obtained with   {\footnotesize CORALIE-07} with a mean signal-to-noise ratio of 52 leading to a mean measurement uncertainty of 3.4~ms$^{-1}$.
 An external systematic error of 5~ms$^{-1}$ was quadratically added to the radial velocity uncertainty before performing 
 the period search and the model adjustment. It took more than 1400 days to realize the importance of this target. Indeed, the first
 5 measurements only showed a quadratic drift that betrayed the presence of a long period companion. We had to wait the sixth 
 measurement, taken in June 2007,  to realize that we missed a planet due to an inadequate temporal sampling ( compared to the period and the phase  of the planet). In the following months,  the presence of a second long period companion, with similar radial velocity amplitude was discovered. 
  It took another one and half year to disentangle the two orbital solutions and to characterize the second planet orbital parameters. 
 
 The planetary system consists of  two massive giant planets with  respective semimajor axis  $a =0.24$~AU and $a =1.92$~AU.
 The first planet is eccentric with $e = 0.469$ and a period of $P = 44.24$~days. It has a minimum mass $m_{b}.\sin{(i)}= 2.12$~M$_{\rm jup}$. The second planet has a much longer period ($P=1008$~days) and a minimum mass  of $m_{c}.\sin{(i)}= 6.56$~M$_{\rm jup}$. Its orbit is slightly eccentric ($e=0.13$) which could betray the presence of interactions between the two massive planets. 
 
\begin{figure}[th!]
\includegraphics[angle=0,width=0.45\textwidth,origin=br]{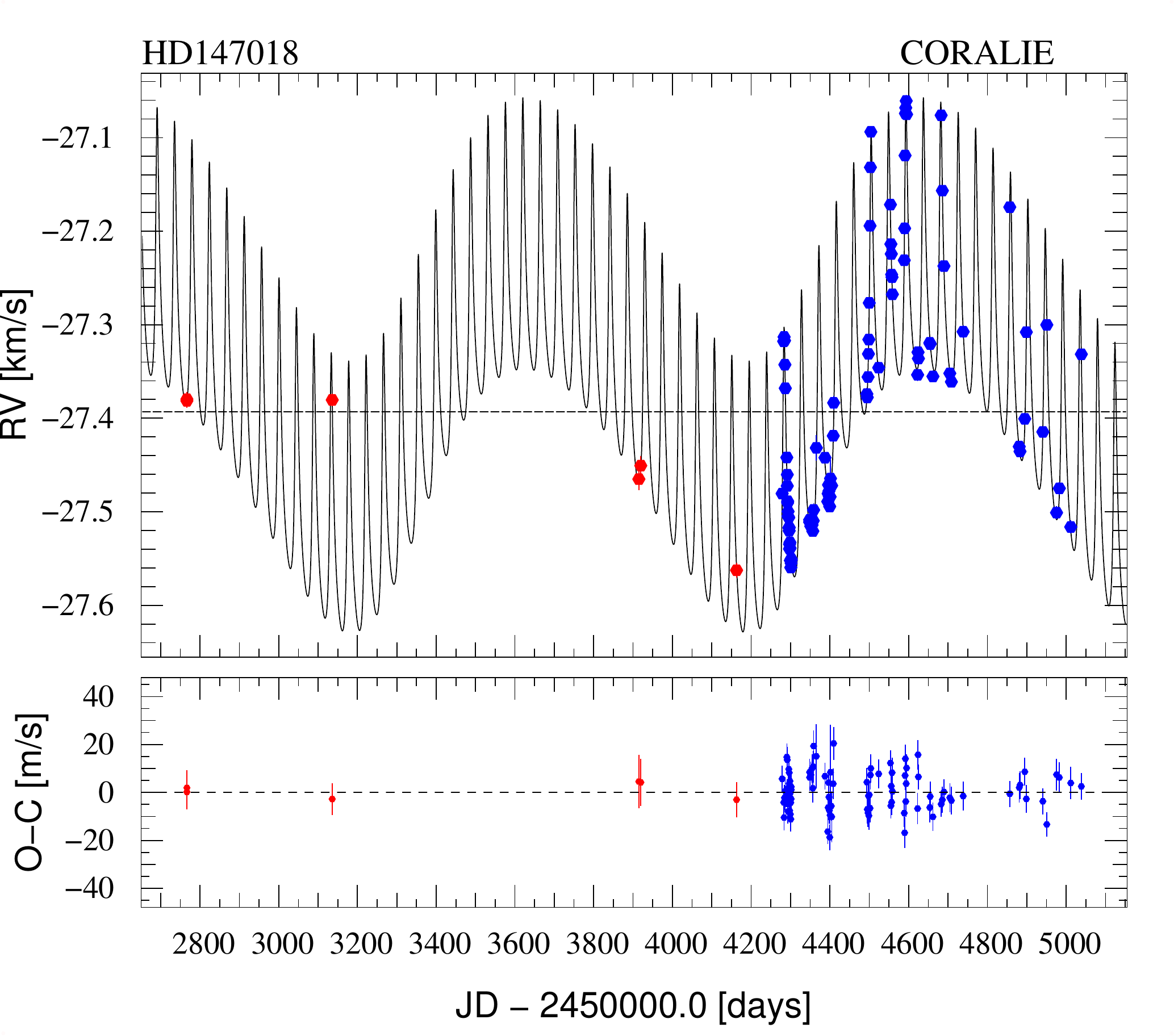} 
\includegraphics[angle=0,width=0.45\textwidth,origin=br]{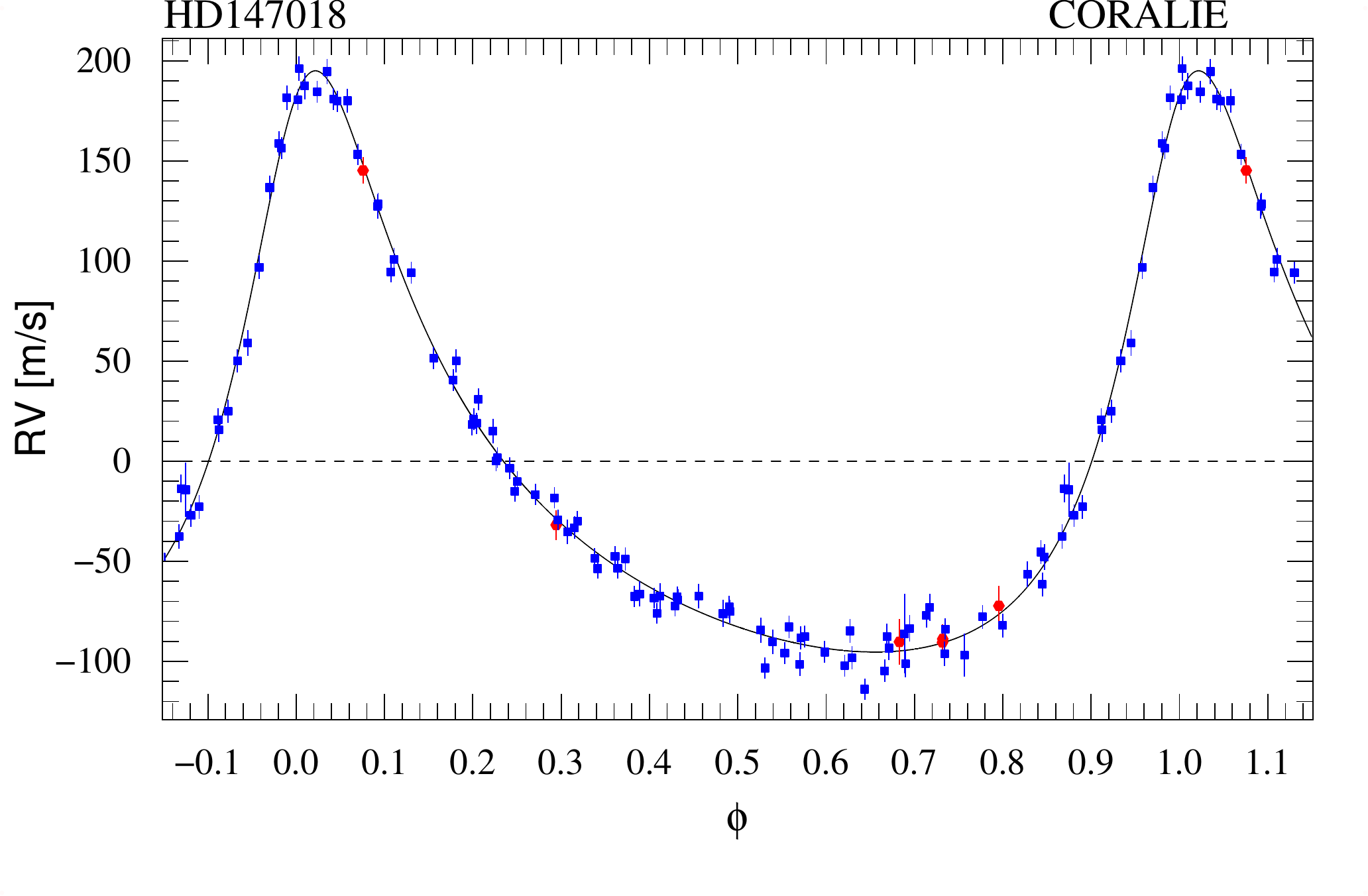} 
 \includegraphics[angle=0,width=0.45\textwidth,origin=br]{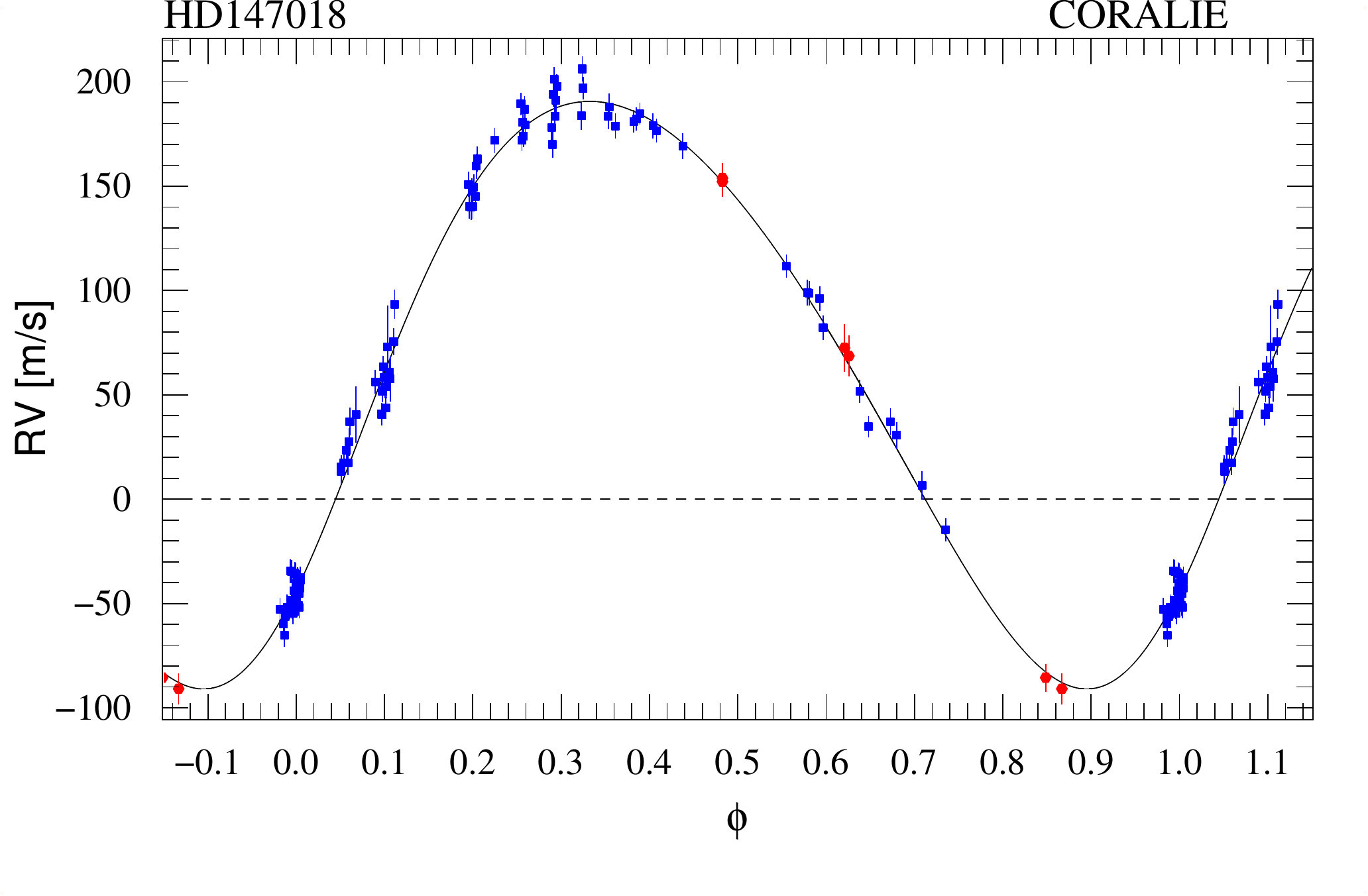}
\caption[]{
\label{fig:rvhd14718}
Radial-velocity measurements of {\footnotesize HD}\,147018 obtained with {\footnotesize CORALIE-98} (red) and  
{\footnotesize CORALIE-07} (blue). The top panel presents the observed radial velocities as a function 
of Julian Date with the best 2-planet-keplerian model (black curve). Residuals show a 7.4~ms$^{-1}$ dispersion.
 The two bottom figures represent the phase folded radial velocities of {\footnotesize HD}\,147018~b (middle)
and of {\footnotesize HD}\,147018~c (bottom).}
\end{figure}

 Figure~\ref{fig:rvhd14718} shows the {\footnotesize CORALIE} 
radial velocities  and the   the adjusted  2 planet-keplerian model.
The residuals to the model show a level  of variation of $\sigma=7.4$~ms$^{-1}$, yielding a reduced $\chi$ of 1.28.
 The orbital elements for  \object{ {\footnotesize HD}\,147018}~b and \object{ {\footnotesize HD}\,147018}~c are listed in Table~\ref{table:orbitalelementshd147018}.
 Error bars were computed using 5000 Monte Carlo  simulations and a confidence interval of 68.3\%.

\begin{table}
\caption{
\label{table:orbitalelementshd147018}
Two planet Keplerian orbital solution
for  \object{ {\footnotesize HD}\,147018} as well as inferred planetary parameters.
 Confidence intervals are computed for a  68.3\% confidence level 
 after 5000 Monte Carlo iterations.
$\Delta T$ is the time interval  between the first and last
measurements, $\chi_{r}$ is the reduced $\chi$, G.o.F. is the Goodness of Fit and
 $\sigma(O-C)$ is the weighted r.m.s. of the residuals
around the derived solution. The Julian Date is expressed as JD$^{*}$=JD-2450 000. 
 {\footnotesize  C98} stands for { \footnotesize CORALIE-98} and  {\footnotesize  C07} for {\footnotesize  CORALIE-07}.}
\begin{center}                                                                   
\begin{tabular}{llcc}                                                           
\hline\hline                                                                     
Planet          &                                & \object{ {\footnotesize HD}\,147018}~b   & \object{ {\footnotesize HD}\,147018}~c         \\
\hline                                                                                                                 
$\gamma_{{\sc C98}}$            &  [kms$^{-1}$]          &\multicolumn{2}{c}{  -27.343 $\pm$ 0.011}        \\
$\gamma_{{\sc C07}}$            &  [kms$^{-1}$]          &\multicolumn{2}{c}{  -27.348 $\pm$ 0.005}        \\\hline                                                                                                                 
$P$                 &  [days]                    &  44.236 $\pm$   0.008      & 1008$\pm$18        \\
$K$                 & [ms$^{-1}$]             &  145.33 $\pm$     1.66       & 141.2$\pm$4.1         \\
$e$                 &                            &   0.4686 $\pm$   0.0081     &  0.133$\pm$0.011         \\
 $\omega$           & [deg]                      &  -24.03$\pm$   1.23         &   -133.1$\pm$6.9        \\
$T_{0}$             &[JD$^{*}$]                     &  4459.49 $\pm$   0.10      &  55301$\pm$22           \\
 \hline                                                                                                                
$a_{1}\, \sin{i}$      & [$10^{-3}$~AU    ]             &  0.5220$\pm$0.0057         &  13.0$\pm$0.6                      \\
$f_{1}(m)$              & [$10^{-9}$~M$_{\odot}$  ]      &   9.70$\pm$0.33           &  287$\pm$30                        \\
 $m_{p}\, \sin{i}$  & [M$_{\rm Jup}$]                 &  2.12$\pm$0.07          &  6.56$\pm$0.32                          \\
  $a$               & [$AU$    ]                    &  0.2388$\pm$0.0039                     &  1.922$\pm$0.039                            \\
\hline                                                                   
N$_{\rm mes}$            &                           &  \multicolumn{2}{c}{  101 }                                   \\
$\Delta T$          & [years]                      &     \multicolumn{2}{c}{  6.22}                            \\
$\chi_{r}$                          &                            & \multicolumn{2}{c}{  1.28$\pm$0.07 }                      \\ 
G.o.F                           &       & \multicolumn{2}{c}{  3.64 }                              \\ 
$\sigma_{(O-C)}$     & [ms$^{-1}$]      & \multicolumn{2}{c}{  7.39  }                              \\   
\hline
\end{tabular}
\end{center}
\end{table}

 \subsection{A long period and massive planet around {\footnotesize HD}\,171238}
\begin{figure}[th!]
   \includegraphics[angle=0,width=0.45\textwidth,origin=br]{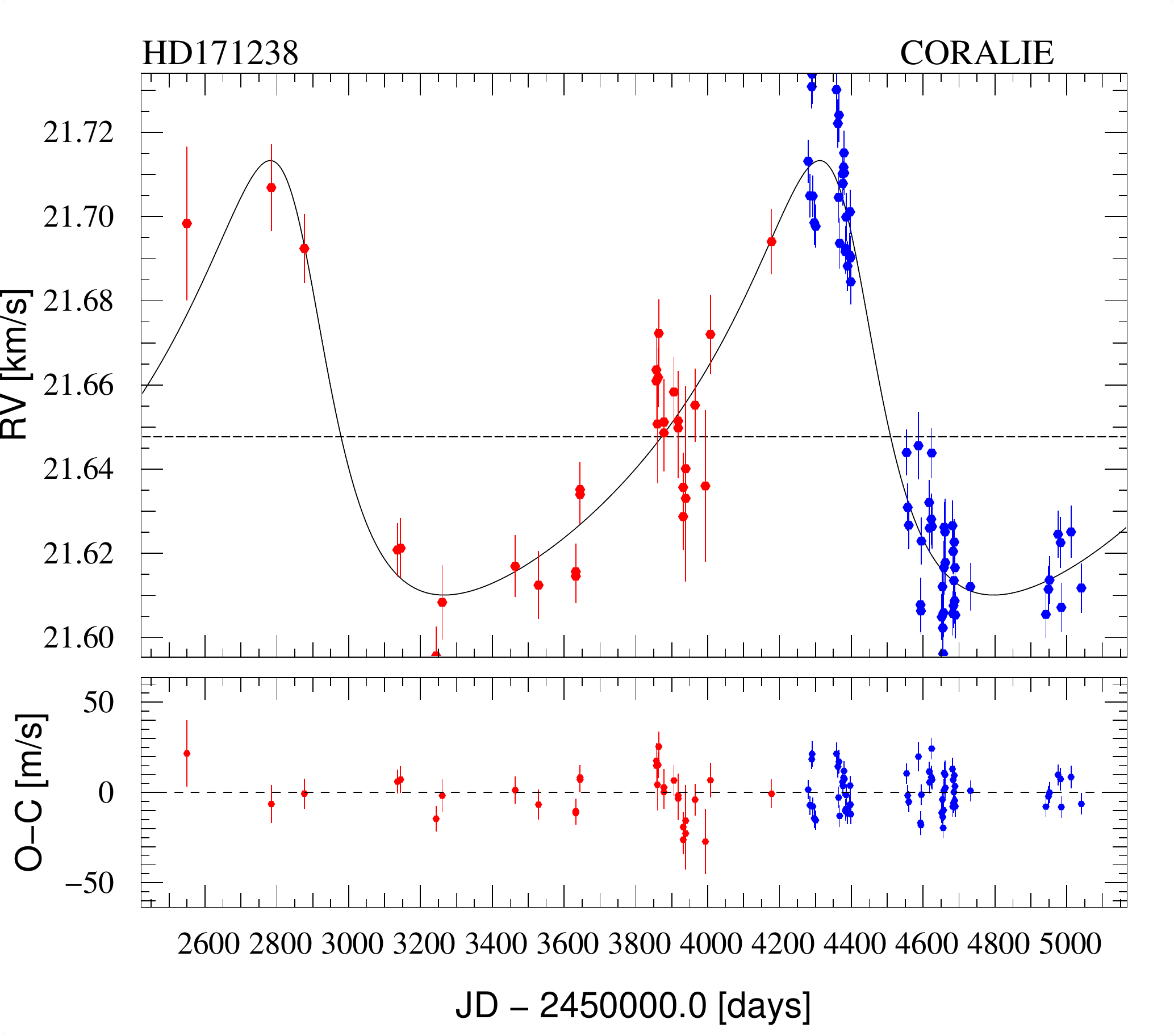} 
   \includegraphics[angle=0,width=0.45\textwidth,origin=br]{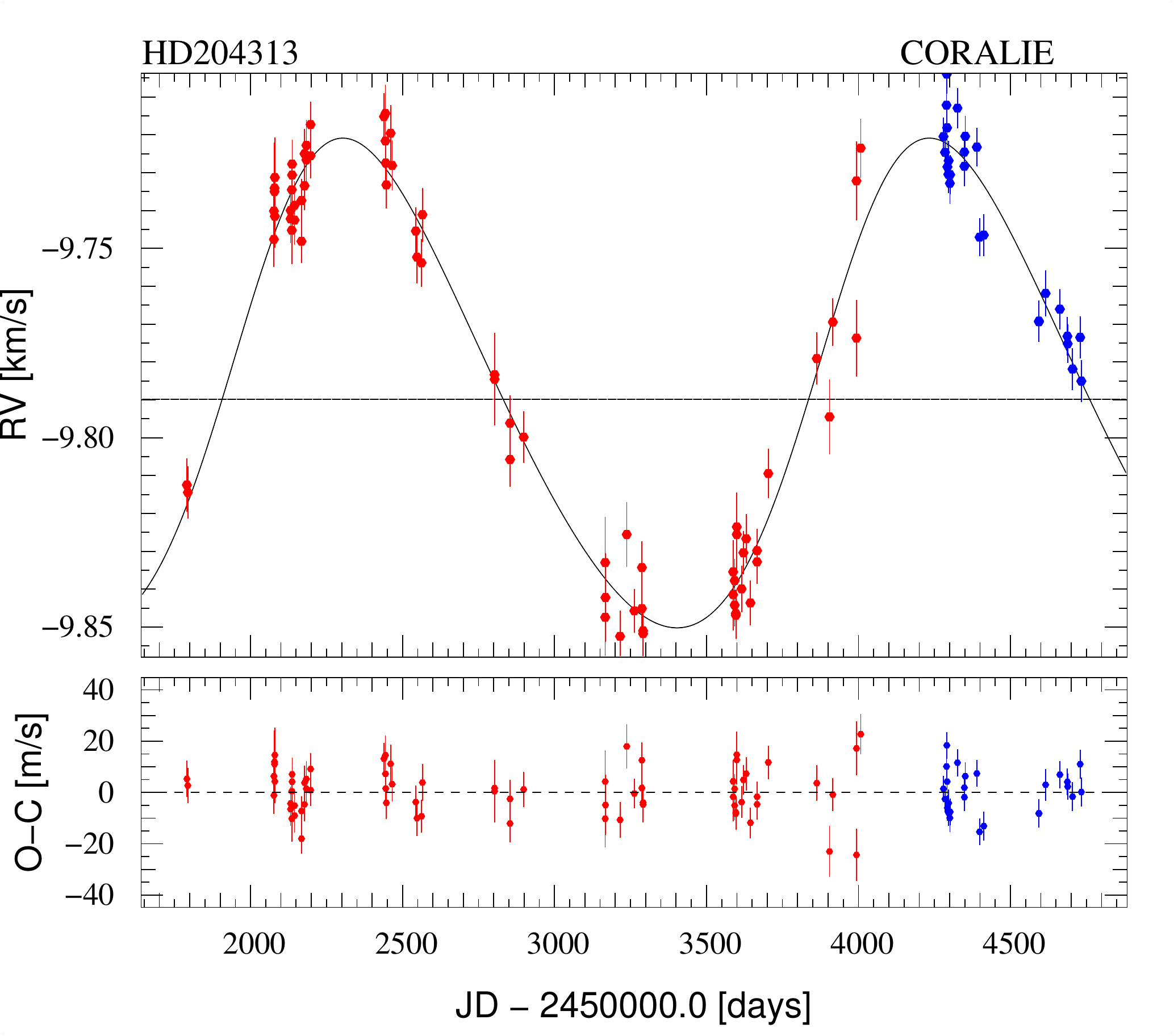} 
\caption[]{
\label{fig:rv-hd171238andhd204313}
The two diagrams represent the observed radial velocities as a function 
of Julian Date obtained with   {\footnotesize CORALIE-98} (red) and  {\footnotesize CORALIE-07} (blue) for {\footnotesize HD}\,171238 and  {\footnotesize HD}\,204313. The best single-planet keplerian model is represented as a black curve and residuals show a dispersion of   10~ms$^{-1}$ for {\footnotesize HD}\,171238  and of 8~ms$^{-1}$ for {\footnotesize HD}\,204313.}
\end{figure}

\begin{figure}[th!]
   \includegraphics[angle=0,width=0.45\textwidth,origin=br]{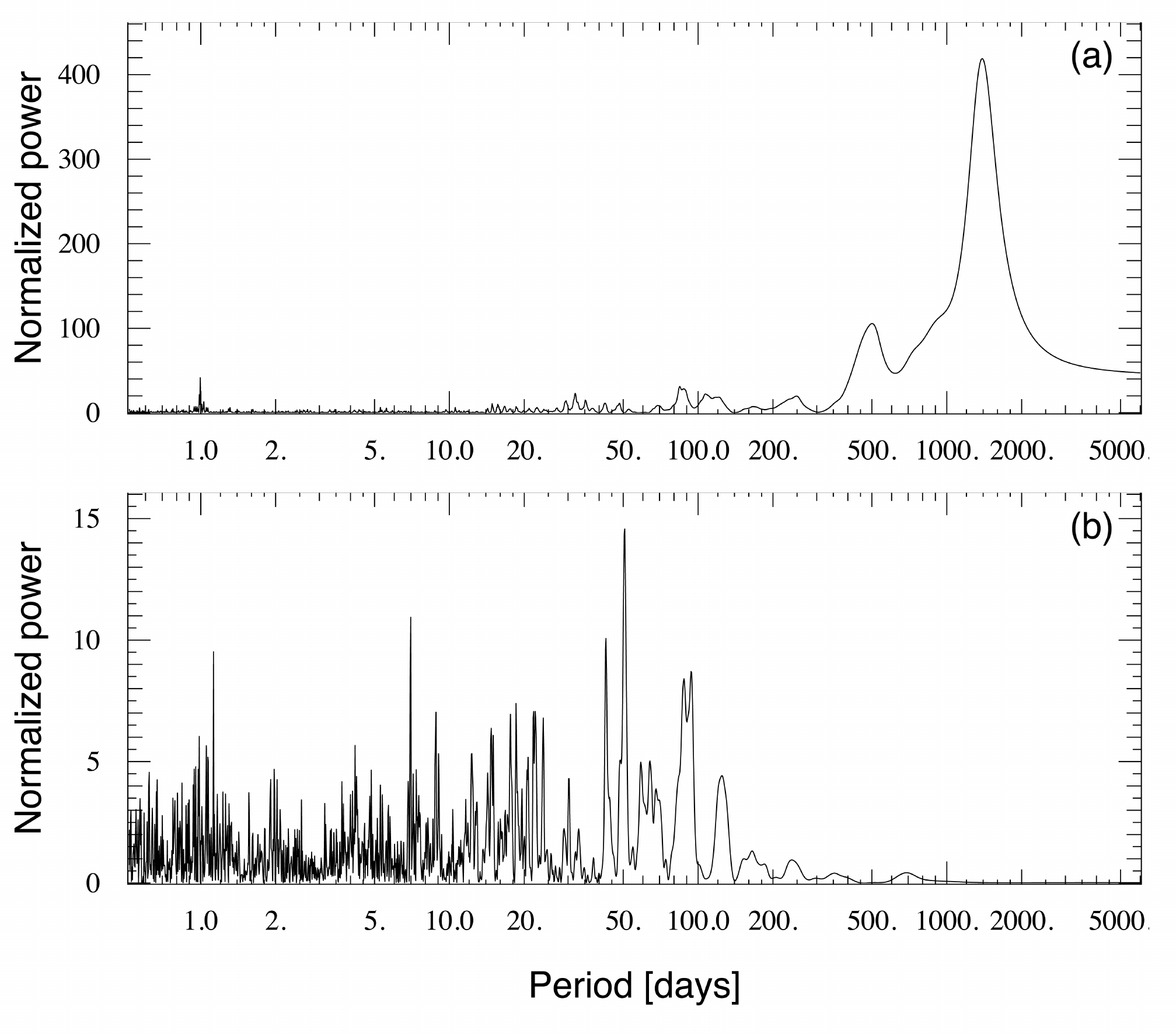} 
\caption[]{
\label{fig:periodogram-hd171238}
Periodogram of  {\footnotesize HD}\,171238 velocities (a) and of the residuals after substraction of the single planet model (b). A significant amount of energy is detected in the residual's close to a period of 50 days which is probably activity induced.}
\end{figure}

 \object{ {\footnotesize HD}\,171238}  has been observed since October 2002. 
 Thirty two radial-velocity  measurements with a typical signal-to-noise ratio of 17 
(per pixel at 550 nm) were obtained with   {\footnotesize CORALIE-98} leading to a mean measurement uncertainty of 6.8~ms$^{-1}$, including photon noise and calibration errors. An additional 65 radial-velocity  measurements  were obtained with   {\footnotesize CORALIE-07} with a mean signal-to-noise ratio of 47 leading to a mean measurement uncertainty of 3.6~ms$^{-1}$.
An instrumental error of 5~ms$^{-1}$ was quadratically added to the radial velocity uncertainty before performing 
 the period search and the model adjustment. A clear signature is identified in the periodogram at $P = 1523$~days
(see Fig. \ref{fig:periodogram-hd171238}) which corresponds to  a massive planet ($m_{b}.\sin{(i)}=2.60$~M$_{\rm jup}$) 
with a semi-major axis of  $a =2.54$~AU and with an eccentric orbit ($e=0.40$).
Figure~\ref{fig:rv-hd171238andhd204313} shows the {\footnotesize CORALIE} 
radial velocities  and the  corresponding best-fit Keplerian model. 
The orbital elements for  \object{ {\footnotesize HD}\,171238}~b are listed in Table~\ref{table:orbitalelementshd171238andhd204313}.  

 The residuals to the adjusted single planet keplerian model are however quite large for {\footnotesize CORALIE} ($\sigma=10$~ms$^{-1}$), 
  yielding a reduced $\chi$ of 1.60. In order to explain such a large dispersion, we have conducted a frequency analysis of the residuals. As can be seen on  Fig. \ref{fig:periodogram-hd171238}, a significant amount of energy is present around 60 and 160 days. However,  no
"realistic" keplerian could be adjusted with such periods, discarding the presence of additional jovian planets. Furthermore, as explained in section \ref{sec:characteristics},
\object{ {\footnotesize HD}\,171238}  is a DY Drac variable star that could vary on time scales of 1 to 120 days.
The most likely explanation for the large radial velocity dispersion is the presence of stellar spots on the surface of the star
which is confirmed by a relatively strong CaII-H re-emission that can be seen in the spectra as illustrated in Fig.\ref{fig:CaIIH-hd171238}
   
  \subsection{A long period and massive planet around  {\footnotesize HD}\,204313}
  \object{ {\footnotesize HD}\,204313}  has been observed   since September 2000.  
  Seventy one radial-velocity  measurements with a typical signal-to-noise ratio of 28 
(per pixel at 550 nm) were obtained with   {\footnotesize CORALIE-98} leading to a mean measurement uncertainty of 5.2~ms$^{-1}$, including photon noise and calibration errors. An additional 26 radial-velocity  measurements  were obtained with   {\footnotesize CORALIE-07} with a mean signal-to-noise ratio of 74 leading to a mean measurement uncertainty of 3.2~ms$^{-1}$. An instrumental error of 5~ms$^{-1}$ was quadratically added to the radial velocity uncertainty before performing 
 the period search and the model adjustment. 
Radial velocity measurements betray the presence of a long period  
($P=1931$~days) and  massive planet ($m_{b}.\sin{(i)}=4.05$~M$_{\rm jup}$) orbiting its parent star at  $a =3.08\,$AU.
The orbital elements for 
\object{ {\footnotesize HD}\,204313}~b are listed in Table~\ref{table:orbitalelementshd171238andhd204313}. 
Figure~\ref{fig:rv-hd171238andhd204313} shows the {\footnotesize CORALIE} 
radial velocities  and the  corresponding best-fit Keplerian model.
 The residuals to the single planet keplerian model show a level
   of variation ($\sigma=8.1$~ms$^{-1}$), yielding a reduced $\chi$ of 1.24.

\begin{table}
\caption{
\label{table:orbitalelementshd171238andhd204313}
Single planet Keplerian orbital solutions
for  \object{ {\footnotesize HD}\,171238} and  for  \object{ {\footnotesize HD}\,204313}
 as well as inferred planetary parameters.
 Confidence intervals are computed for a  68.3\% confidence level 
 after 5000 montecarlo iterations.
$\Delta T$ is the time interval  between the first and last
measurements, $\chi_{r}$ is the reduced $\chi$, G.o.F. is the Goodness of Fit and
 $\sigma(O-C)$ is the weighted r.m.s. of the residuals
around the derived solution. The Julian Date is expressed as JD$^{*}$=JD-2450 000.
C98 stands for { \footnotesize CORALIE-98} and C07 for {CORALIE-07}.}
\begin{center}
\begin{tabular}{llccc}
\hline\hline
Parameters          &                            &      \object{ {\footnotesize HD}\,171238}~b          & \object{ {\footnotesize HD}\,204313}~b              \\
\hline                                                                                                   \\[-2mm]                                                                   
$\gamma_{\sc C98}$            &  [kms$^{-1}$]              &     $21.662\pm0.006$ & $-9.785\pm0.0018$    \\  [1mm]
$\gamma_{\sc C07}$            &  [kms$^{-1}$]         &     $21.641\pm0.002$ &      $-9.762\pm0.0031$   \\
\hline                                                                                       \\[-2mm]                                                                                                            
$P$                 &  [days]                    &     $1523_{-45}^{+40}$   & $1931\pm18$       \\  [1mm]
$K$                 & [ms$^{-1}$]                 &     $52.2\pm1.8$  &   $64.8\pm1.5$        \\  [1mm]
$e$                 &                            &     $0.400_{-0.065}^{+0.061}$  &  $0.131\pm0.023$      \\  [1mm]
 $\omega$           & [deg]                      &     $47.0_{-9.8}^{+9.9}$   &  $-57\pm11$           \\  [1mm]
$T_{0}$             &[JD$^{*}$]                   &      $5062\pm20$   & $ 3989\pm62$       \\[1mm]
 \hline                                                                  
$a_{1}\, \sin{i}$      & [$10^{-3}$~AU    ]         &     $6.68\pm0.32 $             &        $11.40\pm0.31 $      \\
$f_{1}(m)$              & [$10^{-9}$~M$_{\odot}$  ]   &   $17.3\pm2.4$              &    $  53.2\pm4.0   $      \\
$m_{p}\, \sin{i}$  & [M$_{\rm Jup}$]             &  $2.60\pm0.15 $               &  $4.05\pm0.17 $            \\
  $a$               & [AU]                      &  $2.54\pm0.06$                 &        $  3.082\pm0.055 $      \\
                                                                         
\hline                                                                   
N$_{\rm mes}$            &                           &   96             & 97                \\
$\Delta T$        & [years]                    &  6.82              &8.06                 \\
$\chi_{r}$                          &                &  $1.60\pm0.07$ &   $1.24\pm0.07$               \\ 
G.o.F                           &                       &     7.41&        3.09                 \\ 
$\sigma_{(O-C)}$     & [ms$^{-1}$]                &  10.25             & 8.11                \\                        \\   
\hline
\end{tabular}
\end{center}
\end{table}


\section{Concluding discussion\label{sec:discussion}}
We have reported in this paper the detection of four extrasolar planet candidates
 discovered with the {\footnotesize CORALIE} echelle spectrograph  mounted on  the 1.2-m Euler 
 Swiss telescope at La Silla Observatory.\\
 -- {\footnotesize HD}\,147018~b and {\footnotesize HD}\,147018~c are two massive planets part of the same system with respective masses $m_{b}\sin{i}=2.12$~M$_{\rm Jup}$ and $m_{c}\sin{i}$=6.56~M$_{\rm Jup}$. The inner planet has a  44.24 day-period and a large eccentricity ($e=0.46$) while the outer planet has a 1008 day-period with a low eccentricity.\\
 --\object{ {\footnotesize HD}\,171238}~b is a long period and massive planet with an eccentric orbit
 (P=4.14 years, $m\sin{i}=2.60$~M$_{\rm Jup}$, $e=0.40$).\\
 --\object{ {\footnotesize HD}\,204313}~b is a long period and massive planet with a low eccentricity
 (P=5.28 years, $m\sin{i}=4.0$~M$_{\rm Jup}$, $e=0.13$).\\

\begin{figure}[th!]
   \includegraphics[angle=0,width=0.45\textwidth,origin=br]{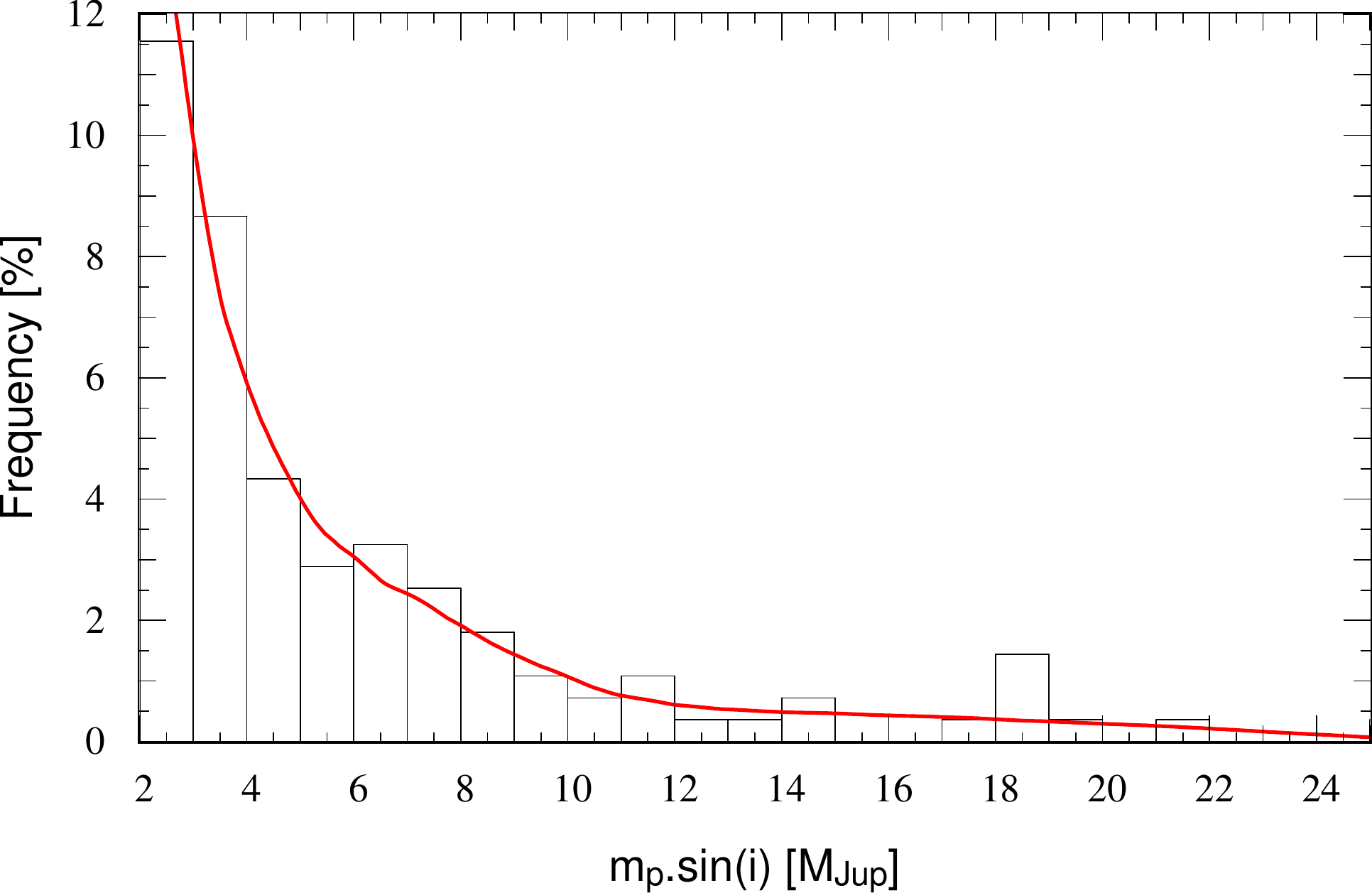}
\caption[]{
\label{fig:histogram_m2sini}
Frequency distribution of the observed $m_{p}.\sin{(i)}$  values for the 117  known exoplanets  with masses larger than
 2~M$_{\odot}$ discovered by radial velocities orbiting G\&K dwarfs. The underlying $m_{p}.\sin{(i)}$  distribution, plotted as a red line, is 
retrieved using a non-parametric approach with an Epanechnikov adaptive Kernel as described in \cite{Jorissen-2001}. }
\end{figure}

It is worth to note that the three parent stars are metal rich, which strengthened the case that massive planets tend to form around metal rich stars
as stated by (\cite{Santos-2001:a,Santos-2005, Fischer-2005}  ).  \\
On the statistical point of view,  the giant planet mass distribution decreases with a power law, as illustrated on 
Fig.~\ref{fig:histogram_m2sini}, with no cut-off or change of distribution morphology up to 25-Jupiter-masses. 
There is no indication in the mass distribution diagram (within 6~AU of the parent star) of the presence of
the low mass tail of a possible brown-dwarf mass distribution. The brown dwarf desert is therefore extremely "dry"
at the lowest masses within 6~AU. However, with only 26 planets candidates with masses larger than 7~M$_{\rm Jup}$,
 one should be carefull not to over-interpret the observations and only a direct measurement of the orbital inclination 
 of each planet candidate will reveal their true nature.

Direct measurement of the orbit inclination of massive planet candidates already produced some results. 
 On one hand, \cite{Bean-2007}, with the HST Fine Guidance sensor, conducted a set of astrometric measurements of
 \object{ {\footnotesize HD}\,33636}   at a fraction of milli-arcsecond accuracy. The authors 
showed that  \object{ {\footnotesize HD}\,33636\,b} is not in the planet domain and is indeed an M dwarf. On the other hand and surprisingly enough, the CoRot space mission 
found  a 21.66-M$_{\rm Jup}$ transiting brown dwarf with a 4.26-day period \cite{Deleuil-2008}.

\begin{figure}[th!]
   \includegraphics[angle=0,width=0.45\textwidth,origin=br]{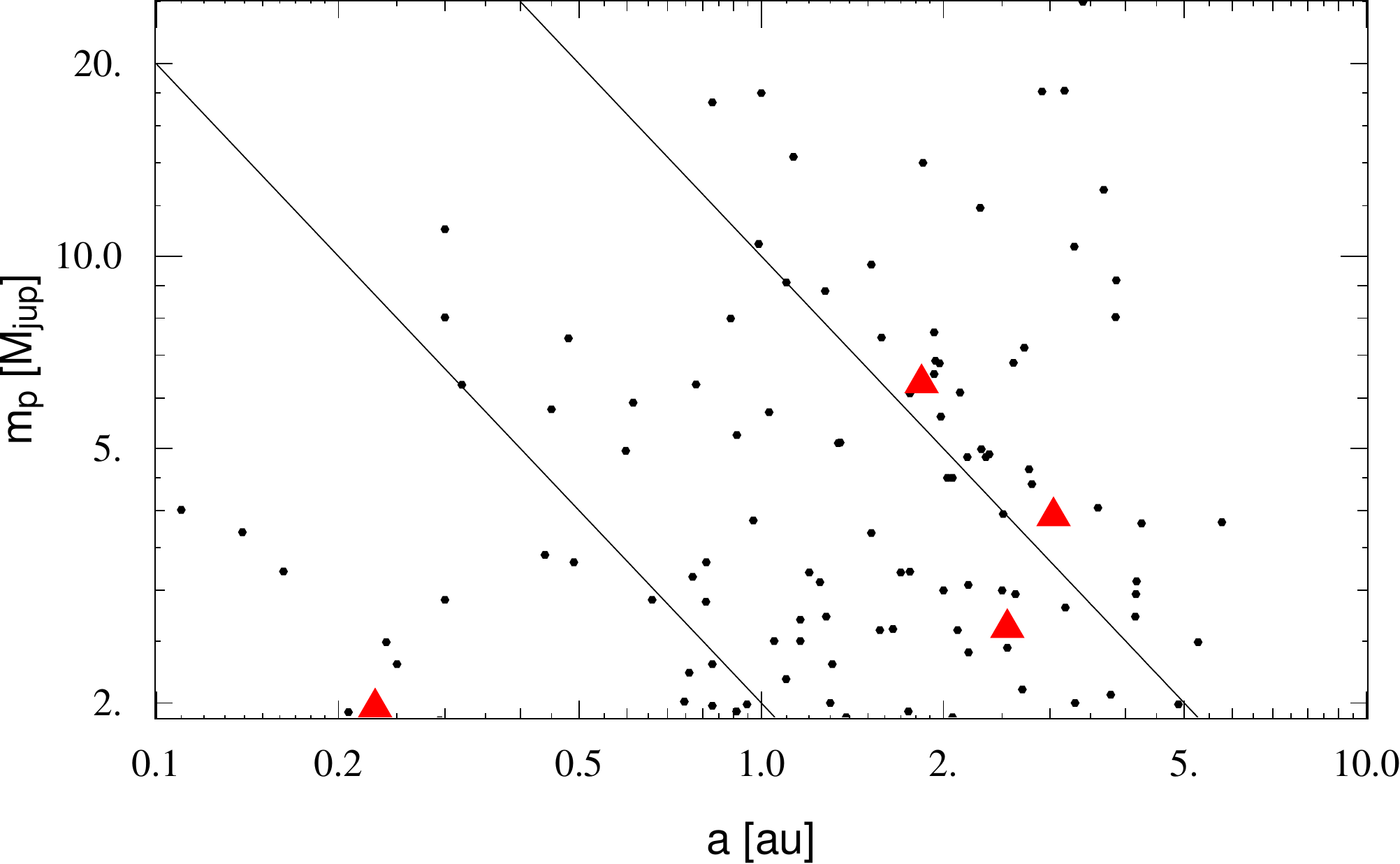}
\caption[]{
\label{fig:mp_au_astrometry}
$m_{p}.\sin{(i)}$ vs.  separation diagram of the known massive exoplanets  orbiting G\&K dwarfs and discovered by radial velocities.
The two lines correspond to astrometric signatures of 50~$\mu$'' and  250~$\mu$'' for a 1~M$_{\odot}$ parent star located 
at 40~pc. The four planets discussed in this paper are represented by  triangles. Three of them will be characterized by {\footnotesize PRIMA} or by {\footnotesize GAIA}.
}
\end{figure}

However, a systematic monitoring of the massive planet candidates will only be possible with the forthcoming dedicated astrometric facilities such as 
PRIMA (start of operation end of 2009) \citep{Launhardt-2008} and GAIA (\citet{Perryman-2001}, launch date end of 2011). Those instruments/telescopes will determine the real mass of all massive planets candidates  as illustrated by figure~\ref{fig:mp_au_astrometry} and provide statistically reliable numbers about the distribution of massive planets and of brown dwarfs within 6~AU.

\begin{acknowledgements}
We thank B. Pernier for participating to {\footnotesize  CORALIE} the observations. We are grateful to the Geneva  Observatory technical staff, in particular to L. Weber, for maintaining the 1.2-m Euler 
Swiss telescope and the {\footnotesize CORALIE} Echelle spectrograph.
 We thank the Swiss National Research  Foundation (FNRS) and the Geneva University for their continuous 
support to our planet search programmes. N.C.S. would like to thank the support from Funda\c{c}\~ao para a Ci\^encia e a Tecnologia, Portugal, through programme Ci\^encia\,2007. Support from the Funda\c{c}\~{a}o para Ci\^{e}ncia e a Tecnologia (Portugal) to P. F. in the form of a scholarship (reference SFRH/BD/21502/2005) is gratefully acknowledged. This 
research has made use of the VizieR catalogue access tool operated at 
CDS, France. 
\end{acknowledgements}

\bibliographystyle{aa}
\bibliography{coralieXVILongPeriodPlanets4AstroPh}	

\end{document}